\documentclass[a4paper]{llncs}
\pdfoutput=1
\usepackage[utf8]{inputenc}
\usepackage{xspace}
\usepackage{amsmath,amssymb,stmaryrd,latexsym}
\usepackage{paralist,url}

\usepackage{comment}
\usepackage{epsfig}
\usepackage{listings}
\usepackage{tikz}
\usetikzlibrary{arrows,backgrounds,positioning,fit,automata,shapes}
\usepackage{semantic}
\usepackage{galois}
\usepackage{algorithmic}
\usepackage{algorithm}
\newcommand{\semb}{[ \! [}
\newcommand{\seme}{] \! ]}

\def\transpose#1{{}^t \! #1}

\def\R{{\mathbb R}}

\def\bfm#1{\protect{\makebox{\boldmath $#1$}}}

\def\x {\bfm{x}}
\def\y {\bfm{y}}

 \newcommand\ForAuthors[1]%          %  temporary remark for the
 {\par\smallskip                     %  authors:
  \begin{center}%                    %
   \fbox%                            %    --------
   {\parbox{0.9\linewidth}%          %    |  #1  |
    {\raggedright\sc--- #1}%         %    --------
   }%                                %
  \end{center}%                      %
  \par\smallskip                     %
 }        

\title{Robustness analysis of finite precision implementations} 
\author{Eric Goubault and Sylvie Putot}
\institute{
  CEA Saclay Nano-INNOV,
  CEA LIST,
  Laboratory for the Modelling and Analysis of Interacting Systems,
  Point Courrier 174,
  91191 Gif sur Yvette CEDEX,
  \{Eric.Goubault,Sylvie.Putot\}@cea.fr
}

\begin{document}

\maketitle

\begin{abstract}
A desirable property of control systems is to be robust to inputs, that is small perturbations of the inputs of a system will 
cause only small perturbations on its outputs. But it is not clear whether this property is maintained at the 
implementation level, when two close inputs can lead to very different execution paths. The problem becomes particularly crucial when 
considering finite precision implementations, where any elementary computation can be affected by a small error. 
In this context, almost every test is potentially unstable, that is, for a given input, the computed (finite precision)
path may differ from the ideal (same computation in real numbers) path. Still, state-of-the-art error analyses 
do not consider this possibility and rely on the stable test hypothesis, that control flows are identical. 
If there is a discontinuity between the treatments in the two branches, that is the conditional block is not 
robust to uncertainties, the error bounds can be unsound.

We propose here a new abstract-interpretation based error analysis of finite precision implementations, relying on the analysis 
of \cite{vmcai11} for rounding error propagation in a given path, but which is now made sound in presence of unstable tests. 
It automatically bounds the discontinuity error coming from the difference between the float and real values when there is a path divergence, 
and introduces a new error term labeled by the test that introduced this potential discontinuity. 
This gives a tractable error analysis,
 implemented in our static analyzer FLUCTUAT: we present results on representative extracts of control programs. % that 
 
\end{abstract}

\section{Introduction}
In the analysis of numerical programs, a recurrent difficulty when we want to assess the influence of finite precision 
on an implementation, is the possibility for a test to be unstable:
when, for a given input, the finite precision control flow can differ from the control flow that would be 
taken by the same execution in real numbers. Not taking this possibility into account may be unsound if the difference of paths leads to 
a discontinuity in the computation,  while taking 
it into account without special care soon leads to large over-approximations. 

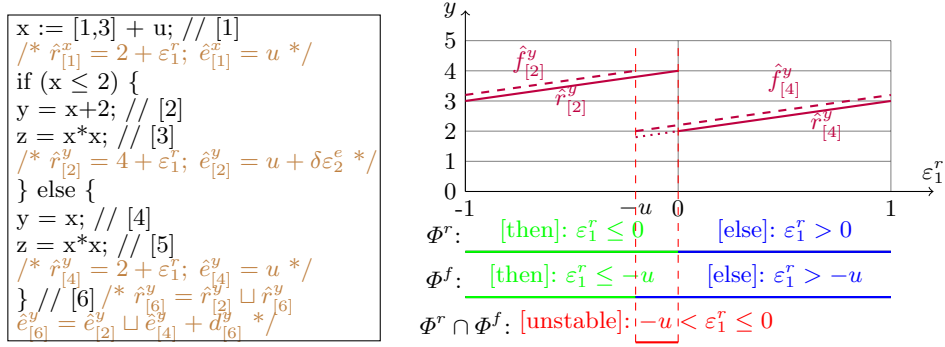
\begin{figure}
\centering
\begin{tikzpicture}[xscale=2.8,yscale=0.4]

\begin{scope}[yscale=0.9]
\draw (-0.75 cm, 6 cm) node[right] {x := [1,3] + u;   // [1]};
\draw (-0.75 cm, 5 cm) node[right,brown] {\small /* $\hat r^x_{[1]} = 2 + \varepsilon_1^r; \; \hat{e}^x_{[1]} = u$ */};
\draw (-0.75 cm, 4 cm) node[right] {if (x $\leq$ 2) \{ };
\draw (-0.75 cm, 3 cm) node[right] { y = x+2;              // [2]};
\draw (-0.75 cm, 2 cm) node[right] { z = x*x;              // [3]};
\draw (-0.75 cm, 1 cm) node[right,brown] {\small /* $\hat r^y_{[2]} = 4 + \varepsilon_1^r; \; \hat{e}^y_{[2]} = u + \delta \varepsilon_2^e$ */};
\draw (-0.75 cm, 0 cm) node[right] { \} else \{ };
\draw (-0.75 cm, -1 cm) node[right] { y = x;                // [4]};
\draw (-0.75 cm, -2 cm) node[right] {  z = x*x;             // [5]};
\draw (-0.75 cm, -3 cm) node[right,brown] {\small /* $\hat r^y_{[4]} =  2 + \varepsilon_1^r; \; \hat{e}^y_{[4]} = u $ */};
\draw (-0.75 cm, -4 cm) node[right] {  \}                   // [6]};
\draw (-0.75 cm, -4 cm) node[right,brown] {\hspace*{1.cm}  \small /* $\hat r^y_{[6]} = \hat{r}^y_{[2]} \sqcup \hat{r}^y_{[6]}$ };
\draw (-0.75 cm, -5 cm) node[right,brown] {\small    $\hat e^y_{[6]} = \hat{e}^y_{[2]} \sqcup \hat{e}^y_{[4]} + d^y_{[6]}$ */ };
\draw (-0.75,-5.5) -- (1.,-5.5) -- (1.,6.5) -- (-0.75,6.5) --cycle;
\end{scope}

\begin{scope}[xshift=0.4cm]
\draw[xstep=1cm,ystep=1cm,gray,very thin] (1,0) grid (3,5); 
\draw[->] (1,0) -- (3.2,0) node[anchor=south] {$\varepsilon_1^r$};
\draw (1 cm,0 cm)  node[below] {-1};
\draw (2 cm,0 cm)  node[below] {0};
\draw (1.8 cm,0 cm)  node[below] {$-u$};
\draw (3 cm,0 cm)  node[below] {1};
\draw[->] (1,0) -- (1,6) node[anchor=east] {$y$};
\foreach \y/\ytext in {0,1,2,3,4,5} 
\draw (1 cm,\y cm) node[left] {$\ytext$};

\draw [domain=2:3,purple,thick] plot (\x,{\x});
\draw [domain=1:2,purple,thick] plot (\x,{\x+2});
\draw [domain=1.8:2,purple,thick,dotted] plot (\x,{\x});
\draw [domain=1:1.8,purple,thick,dashed] plot (\x,{\x+2.2});
\draw [domain=1.8:3,purple,thick,dashed] plot (\x,{\x+0.2});

\draw (0.9 cm,-2 cm) node[anchor=south] {$\Phi^r$:}; %$\varepsilon_1^r$ for real value control flow};
\draw[thick] (1,-2) -- (3,-2);
\draw (0.9 cm,-3.5 cm) node[anchor=south] {$\Phi^f$:}; %$\varepsilon_1^r$ for float value control flow};
\draw[thick] (1,-3.5) -- (3,-3.5);

\draw[thick,green] (1,-2) -- (2,-2);
\draw[green] (1.5 cm,-2 cm) node[anchor=south] {[then]: $\varepsilon_1^r \leq 0$};

\draw[thick,blue] (2,-2) -- (3,-2);
\draw[blue] (2.5 cm,-2 cm) node[anchor=south] {[else]: $\varepsilon_1^r > 0$};

\draw[thick,green] (1,-3.5) -- (1.8,-3.5);
\draw[green] (1.5 cm,-3.5 cm) node[anchor=south] {[then]: $\varepsilon_1^r \leq -u$};

\draw[thick,blue] (1.8,-3.5) -- (3,-3.5);
\draw[blue] (2.5 cm,-3.5 cm) node[anchor=south] {[else]: $\varepsilon_1^r > -u$};

\draw[red,dashed] (1.8,-5) --  (1.8,5);
\draw[red,dashed] (2,-5) --  (2,5); 

\draw (1 cm,-5 cm) node[anchor=south] {$\Phi^r \cap \Phi^f$:}; % $\varepsilon_1^r$ for unstable test};
\draw[thick,red] (1.8,-5) -- (2,-5); 
%\draw[red] (1.975,3.75) ellipse (0.2 and 1);
\draw[red] (1.85 cm, -5 cm) node[anchor=south] {[unstable]: $ -u < \varepsilon_1^r \leq 0$};

\draw[purple] (1.5 cm, 2.2 cm) node[anchor=south] {$\hat r^y_{[2]}$};
\draw[purple] (1.3 cm, 3.4 cm) node[anchor=south] {$\hat f^y_{[2]}$};
\draw[purple] (2.5 cm, 2.8 cm) node[anchor=south] {$\hat f^y_{[4]}$};
\draw[purple] (2.7 cm, 1.4 cm) node[anchor=south] {$\hat r^y_{[4]}$};
\end{scope}

\end{tikzpicture}
\caption{Running example} \label{lst::ex1}
\end{figure}

And when considering programs that compute with approximations of real numbers, potentially unstable tests lie everywhere: 
we want to automatically characterize conditional blocks that perform a continuous treatment of inputs, and are thus 
robust, and those that do not. 
This unstable test problem is thus closely related to the notion of continuity/discontinuity in programs, first introduced in
\cite{DBLP:conf/issta/Hamlet02}. 
Basically, a program is continuous if, when its inputs are slightly perturbed, its output is also
only slightly perturbed, very similarly to the concept of a continuous function.  
Discontinuity in itself can be a symptom of a major bug in some critical systems, such as
the one reported in \cite{Bushnell12}, where a 
F22 Raptor military aircraft almost crashed after crossing the international date line in 2007, due to a discontinuity in the treatment of dates.
Consider the toy program presented on the left hand side of Figure \ref{lst::ex1}, where input $x$ takes its real value in 
$[1,3]$, with an initial error $0 < u << 1$, that can come either from previous finite precision computations, 
or from any uncertainty on the input such as sensor imperfection.
The test is potentially unstable: for instance, if the real value of $x$ at control point [1] is $r^x_{[1]}=2$, then its floating-point value is 
$f^x_{[1]}=2+u$. 
Thus the execution in real numbers would take the \texttt{then} branch and lead at control point [2] to $r^y_{[2]} = r^x_{[1]}+2=4$, whereas the 
floating-point execution would take the  \texttt{else} branch and lead to $f^y_{[4]}=f^x_{[1]}=2+u$.
The test is not only unstable, but also introduces a discontinuity around the test condition $(x == 2)$. %: a small variation on $x$ around $x==2$ 
%leads to a large variation of the output $y$. 
Indeed, for $r^x_{[1]}=2$, there is an error due to discontinuity of $f^y_{[4]}-r^y_{[2]}=-2+u$. 
Of course, the computation of $z$ around the test condition is continuous. \\

In the rest of the paper, we propose a new analysis, that enhances earlier work by the authors~\cite{vmcai11}, by computing and propagating 
bounds on those discontinuity errors. 
This previous work characterized the computation error due to 
the implementation in finite precision, by comparing the computations in real-numbers with the same computations 
in the floating-point semantics, relying on the stable test assumption: the floating-point number
control flow does not diverge from the real number control flow. In its implementation in FLUCTUAT~\cite{fmics2009}, in the case when the analysis 
determined a test could be unstable, it issued a warning, and the comparison between the two semantics could be unsound. 
This issue, and the stable test assumption, appear in all other (static or dynamic) existing analyzes of numerical error propagation; 
the expression unstable test is actually taken from CADNA \cite{CADNA}, a stochastic arithmetic instrumentation of programs, to assert their numerical quality.
In Hoare provers dealing with both real number and floating-point number semantics, e.g. \cite{BoldoFilliatre07}
this issue has to be sorted out by the user, through suitable assertions and lemmas. 

Here as in previous work, we rely on the relational abstractions
of real number and floating numbers semantics using affine sets (concretized as zonotopes)~\cite{arxiv08,arxiv09,cav09,cav10,vmcai11}. 
But we now also, using these abstractions, compute and solve constraints on inputs such that the execution potentially leads to unstable tests, 
and thus accurately bound the discontinuity errors, computed as the difference of the floating-point value in one branch and the real value in another, 
when the test distinguishing these two branches can be unstable.

Let us exemplify and illustrate this analysis on the program from Figure \ref{lst::ex1}. 
%\vspace*{-0.2cm}
%\vspace*{-0.1cm}
The real value of input \texttt{x} will be abstracted by the affine form $\hat r^x_{[1]}=2 + \varepsilon_1^r$, 
where $\varepsilon_1^r$ is a symbolic variable with values in $[-1,1]$. Its error is $\hat{e}^x_{[1]}=u$ and its 
finite precision value is  $\hat{f}^x_{[1]}=\hat{r}^x_{[1]} + \hat{e}^x_{[1]}= 2 + \varepsilon_1^r +u$.
Note the functional abstraction: affine forms represent a function from inputs to variable values. We will use this to interpret tests, and in particular to 
compute unstable tests conditions. 
For instance, the condition for the execution in real numbers to take the \texttt{then} branch is here $2 + \varepsilon_1^r \leq 2$,
that is $\varepsilon_1^r \leq 0$. Now, the condition for the execution in finite precision to take the \texttt{else} branch is $\hat f^x_{[1]}>2$, that is 
$2 + \varepsilon_1^r +u > 2$, which is equivalent to $\varepsilon_1^r > -u$.
Thus, the unstable test condition being that for the same input - or equivalently here the same value of $\varepsilon_1^r$ - the real and float control flow are different, 
this amounts to intersecting these two conditions on $\varepsilon_1^r$, 
and yields $-u < \varepsilon_1^r \leq 0$. These constraints are illustrated on Figure \ref{lst::ex1}, with $u=0.2$: $\Phi_r$ denotes the constraints on the real value, 
$\Phi_f$, the constraints on the finite precision value, and $\Phi^r \cap \Phi^f$, the unstable test condition.
For the other possibility for an unstable test, that is the execution in real numbers takes the \texttt{else} branch while the float
execution takes the \texttt{then} branch, the constraints are $\varepsilon_1^r < 0$ and $\varepsilon_1^r \leq -u$, which are incompatible. This possibility 
is thus excluded.
 We will see later that these constraints allow us in general to 
refine the bounds on the discontinuity error, but they are also useful to characterize the set of inputs that can lead to unstable test: $-u < \varepsilon_1^r \leq 0$
corresponds to $2-u < r^x < 2$. 

Take now variable \texttt{y}. In the \texttt{then} branch, its real value is $\hat r^y_{[2]}= \hat{r}^x_{[1]} + 2 = 4 + \varepsilon_1^r$, the error $\hat e^y_{[2]}= \hat{e}^x_{[1]} + \delta \varepsilon_2^e$, 
where $\delta$ is the bound on the elementary rounding error on \texttt{y}, due to the addition,
we deduce $\hat{f}^y_{[2]}= \hat{r}^y_{[2]} + \hat{e}^y_{[2]}$. In the \texttt{else} branch,  the real value is $\hat r^y_{[4]}= \hat{r}^x_{[1]} = 2 + \varepsilon_1^r$, 
the error $\hat e^y_{[4]}= \hat{e}^x_{[1]}$, and we deduce $\hat f^y_{[4]}= \hat{r}^y_{[4]} + \hat{e}^y_{[4]}$. 
In Figure \ref{lst::ex1}, we represent in solid lines the real value of $y$ and in dashed lines its finite precision value.
With the previous analysis~\cite{vmcai11} that makes the stable test assumption, % propagates errors in each path and simply joins values and errors when path are joined, 
%joining branches at control point [6], 
we compute when joining branches  at control point [6], 
$\hat r^y_{[6]}=\hat{r}^y_{[2]} \sqcup \hat{r}^y_{[4]} = 3 + \varepsilon_6^r \in [2,4]$ with new noise symbol $\varepsilon_6^r$ (note that we will not detail here the upper bound operator
on affine forms, discussed in e.g. \cite{arxiv09,vmcai11,modular}), $\hat e^y_{[6]} = \hat{e}^y_{[2]} \sqcup \hat{e}^y_{[4]} = u  + \delta \varepsilon_2^e \in [u-\delta,u+\delta]$, 
and $\hat f^y_{[6]}=\hat{r}^y_{[6]}+\hat{e}^y_{[6]} = 3 + u + \varepsilon_6^r + \delta \varepsilon_2^e$. This is sound for the real and float values 
$\hat r^y_{[6]}$ and $\hat f^y_{[6]}$, but unsound for the error because of the possibility of an unstable test. 
Our new analysis, when joining branches, also computes bounds for  %$\hat f^y_{[4]}-\hat r^y_{[2]}=2 + \varepsilon_1^r+u-(4 + \varepsilon_1^r)=-2+u$, 
$\hat r^y_{[4]}-\hat{r}^y_{[2]}=2 + \varepsilon_1^r-(4 + \varepsilon_1^r)=-2$
under the unstable test condition  $-u < \varepsilon_1^r \leq 0$ (or $2-u < \hat r^x < 2$): a new discontinuity term is added and
the error is now $\hat e^y_{[6]} + d^y_{[6]}$ where $d^y_{[6]} = -2 \chi_{[-u,0]}(\varepsilon_1)$ and $\chi_{[a,b]}(x)$ equals 1 if $x$ is in $[a,b]$ and 0 otherwise.

\paragraph{Related work}

In \cite{CGL10}, the authors introduce a continuity analysis of programs.
This approach is pursued in particular in 
\cite{DBLP:conf/sigsoft/ChaudhuriGLN11,CGL12}, where several refinements of the
notion of continuity or robustness of programs are proposed, another one being introduced
in  \cite{DBLP:conf/rtss/MajumdarS09}. These notions are discussed in \cite{Gazeau},
in which an interactive proof scheme for proving a general form 
of robustness is discussed.
In \cite{DBLP:conf/rtss/MajumdarS09}, 
the algorithm proposed by the authors symbolically traverses program paths
and collects constraints on input and output variables. Then for each pair of program
paths, the algorithm determines values of input variables that cause the program to follow
these two paths and for which the difference in values of the output variable is
maximized. We use one of their examples (transmission shift, Section \ref{experiments}), and show that we reach similar conclusions. 
One difference between the approaches is that we give extra information concerning the finite precision
flow divergence with respect to the real number control flow, potentially exhibiting flawed behaviors. 
Also, their path-sensitive analysis can exhibit witnesses for worst discontinuity errors, but
at the expense of a much bigger combinatorial complexity. Actually, we will show that our unstable test constraints 
also allow us to provide indication on the inputs leading to discontinuity errors. 

Robustness has also been discussed in the
context of  synthesis and validation of control systems, in 
\cite{DBLP:journals/corr/abs-1108-3540,DBLP:conf/emsoft/TabuadaBCSM12}. 
The formalization is based on automata theoretic methods, providing a 
convenient definition of a metric between B\"uchi automata. % mais pas vraiment d'algo
Indeed, robustness has long been central in numerical mathematics, in particular
in control theory. The field of robust control is actually concerned in proving
stability of controlled systems where parameters are only known in range. 
A notion which is similar to the one of
\cite{DBLP:conf/emsoft/TabuadaBCSM12}, but in the realm of real numbers and
control of ordinary differential equations, is the input-output stability/continuity 
in control systems as discussed in \cite{Sontag}.

This problematic is also of primary importance in
computational geometry, see for instance \cite{Shewchuk} for a survey on the use
of ``robust geometric predicates''. Nevertheless,
the aim pursued is different from ours: we are mostly interested in critical embedded
software, where the limited resources generally prevent the use of complicated, refined
arithmetic algorithms. 

\paragraph{Contents}

Our main contribution is a tractable analysis that generalizes both the abstract domain
of \cite{vmcai11} and the continuity or robustness analyses: it ensures the finite precision error analysis 
is now sound even in the presence of unstable tests, by computing and propagating discontinuity error bounds for these tests. 

We first review in Section \ref{zonotopicreal} the basics of the relational analysis based on affine forms for the abstraction of real number semantics
necessary to understand this robustness analysis presented here. We then introduce in Section \ref{abstraction} our new abstract domain, 
based on an abstraction similar
to that of \cite{vmcai11}, but refined to take care of unstable tests properly. 
We present in Section \ref{sec::technical} some refinements that are useful for reaching more
accurate results, but are not central to understand the principles of the analysis.
We conclude with some experiments using our implementation of this abstraction in our static analyzer
FLUCTUAT.

\section{Preliminaries: affine sets for real valued analysis}

We recall here the key notions on the abstract domains based on affine sets for the analysis of real value of program variables
that will be needed in Sections \ref{abstraction} and \ref{sec::technical} for our robustness analysis. We refer to \cite{sas06,arxiv08,arxiv09,cav09,cav10} for more details.  

\vspace*{-0.2cm}
\subsubsection{From affine arithmetic to affine sets}

\label{zonotopicreal}

% HOP

Affine arithmetic %~\cite{com-sto-93-aa} 
is a more accurate extension of interval arithmetic, 
that takes into account affine correlations between variables. 
An {\em affine form} is a formal sum over a set of {\em noise symbols} 
$\varepsilon_i$ \[ \hat{x} \,\overset{\textup{def}}{=}\, \alpha^x_0 + \sum_{i=1}^n \alpha^x_i \varepsilon_i,\]
with $\alpha^x_i \in \R$ for all $i$. %Let $\RAff$ denote the set of such affine forms. 
Each noise symbol $\varepsilon_i$ stands for an independent component of the total
uncertainty on the quantity $\hat x$, its value is unknown but bounded in [-1,1]; 
the corresponding coefficient $\alpha^x_i$ is a known real value, which gives 
the magnitude of that component. The same noise symbol can be 
shared by several quantities, indicating correlations among them. These noise 
symbols can not only model uncertainty in data or parameters,
but also uncertainty coming from computation.
The values that a variable $x$ defined by an affine form $\hat x$ can take is in the range
$ \gamma(\hat x) = \left[ \alpha^x_0 - \sum_{i=1}^n |\alpha^x_i| , \alpha^x_0 + \sum_{i=1}^n |\alpha^x_i| \right].$

%\noindent {\em \bf Assignment}
The assignment of a variable $x$ whose value is given in a range $[a,b]$,
is defined as a centered form using a fresh noise symbol $\varepsilon_{n+1} \in [-1,1]$, 
which indicates unknown dependency to other variables:
$\hat x = \frac{(a+b)}{2} + \frac{(b-a)}{2} \, \varepsilon_{n+1}$.

%\noindent {\em \bf Affine  operations}
The result of linear operations on affine forms is an affine form, and is thus interpreted exactly. For two affine forms 
$\hat x$ and $\hat y$, and a real number $\lambda$, we have
%\vspace{-.25cm}
$ \lambda \hat x + \hat{y} = (\lambda \alpha^x_0 + \alpha^y_0) + \sum_{i=1}^n(\lambda \alpha^x_i + \alpha^y_i) \varepsilon_i.$
For non affine operations, we select an approximate linear resulting form, and bounds for 
the error committed using this approximate form are computed, 
that are used to add a new noise term to the linear form. %We refer the reader to \cite{sas06,arxiv09} for more details.

As a matter of fact, the new noise symbols introduced in these linearization processes, were
given different names in \cite{arxiv09,vmcai11}: the $\eta_j$ symbols. Although they play a
slightly different role than that of $\varepsilon_i$ symbols, for sake of notational simplicity,
we will only give formulas in what follows, using the same $\varepsilon_i$ symbols for both
types of symbols. 
%The main property of the zonotopic abstract domain is that it is a functional
%abstraction, meaning that it abstracts 
The values of the variables at a given control point
as a linearized function of the values of the inputs of the program, that we generally identify with a prefix
of the $\varepsilon_i$ vector. %This means that the abstract value gives a linearisation
%of the input-output function, computed by a program, up to some new 
The uncertainties, due
to the abstraction of non-linear features such as the join and the multiplication  will be abstracted on 
a suffix of the $\varepsilon_i$ vector - previously the $\eta_j$ symbols. 

In what follows, we use the matrix notations of \cite{arxiv09} to handle affine sets, that is tuples of affine forms.
We note ${\cal M}(n,p)$ the space of matrices with $n$ lines and $p$ columns of real coefficients.
A tuple of affine forms expressing the set of values taken by $p$ variables over $n$ noise symbols 
$\varepsilon_i, \; 1 \leq i \leq n$, can be represented by a matrix $A \in {\cal M}(n+1,p)$. 

\vspace*{-0.2cm}
\subsubsection{Constrained affine sets}
\label{constrainedzonotopes}

As described in \cite{cav10}, we interpret tests by adding some constraints on the $\varepsilon_i$ noise symbols, 
instead of having them vary freely into [-1,1]: we restrain ourselves to executions (or inputs) that can take the considered branch. 
We can then abstract these constraints in any abstract domain, the simplest being intervals, but we will see 
than we actually need (sub-)polyhedric abstractions to accurately handle unstable tests. We note $\cal A$ for this 
abstract domain, and use $\gamma: {\cal A} \rightarrow \wp(\R^n)$ for the concretisation operator, and
$\alpha: \wp(\R^n) \rightarrow {\cal A}$ for some ``abstraction'' operator, not necessarily the best one (as in polyhedra,
this does not exist): we only need to be able to get an abstract value from a set of concrete values, such that
$X \subseteq \gamma \circ \alpha(X)$. %The order on $\cal A$ will be denoted by $\leq$ in the sequel. %, a fairly overloaded
%symbol, but which should be clear according to context.

This means that abstract values $X$ are now composed of a zonotope identified with its
matrix $R^X \in {\cal M}(n+1,p)$, together with an abstraction $\Phi^X$ of the constraints on the noise symbols, 
$X =(R^X,\Phi^X)$.
%For instance, we can associate to
% function $\Phi^X$, which, to 
% each $\varepsilon_i$ an interval constraint (within -1 and 1). 
The concretisation of such constrained zonotopes or affine sets is 
$\gamma(X)=\left\{\transpose{C^X} \epsilon \mid \epsilon \in \gamma(\Phi^X) \right\}.$
For $\Phi \in {\cal A}$, and $\hat x$ an affine form, we note
$\Phi(\hat x)$ the interval $[J^{-},J^{+}]$ with $J^{-}$ and $J^{+}$ given by the linear programs
$J^{-}=\inf_{\varepsilon \in \gamma(\Phi)} \hat{x}(\varepsilon)$ and $J^{+}=\sup_{\varepsilon \in \gamma(\Phi)} \hat{x}(\varepsilon)$.

\begin{example}
For instance on the running example, 
starting with program variable $x$ in $[1,3]$, we associate the abstract value $X$ with
$R^X=(2\mbox{ } 1)$, i.e. $\hat x = 2 + \varepsilon_1$, and $\gamma(\Phi^X)=\gamma(\varepsilon_1)=[-1,1]$. The interpretation
of the test \texttt{if (x<=2)} in the \texttt{then} branch %does not change the
%zonotopic part of the abstract value (the affine form), but $x\leq 2$ 
is translated into constraint $\varepsilon_1 \leq 0$, thus $\gamma({\Phi^X})=[-1,0]$. Then, the interval concretisation of $\hat x$ is $\gamma(\hat x)=[2-1,2]=[1,2]$.
\end{example}

\subsubsection{Transfer functions for arithmetic expressions}

Naturally, the transfer functions described in the unconstrained case are still correct when we have additional constraints on the 
noise symbols; but for the non linear operations such as the multiplication, the constraints can be used to refine the result by computing 
more accurate bounds on the non affine part which is over-approximated by a new noise term, solving with a guaranteed
linear solver\footnote{For an interval domain for the constraints on noise symbols, a much more straightforward 
computation can be made, of course.} the linear programming problems $\sup_{\epsilon \in \gamma(\Phi^X)} \varepsilon$ (resp. $\inf$).
 Transfer functions are described, respectively in the 
unconstrained and constrained cases in  \cite{arxiv09} and \cite{cav10}, and will not 
be detailed here, except in the example below.

\begin{example}
\label{runex2}
Consider the computation \texttt{z=x*x} at control point $3$ in the \texttt{then} branch of the running example
(Figure \ref{lst::ex1}). If computed as in the unconstrained case, 
we write $\hat z_{[3]} = (2 + \varepsilon_1) (2 + \varepsilon_1) = 4 + 4 \varepsilon_1 + (\varepsilon_1)^2$, which, using the fact that $(\varepsilon_1)^2$
is in [0,1], can be linearized using a new noise symbol by 
$\hat z_{[3]} = 4.5 + 4 \varepsilon_1 + 0.5 \varepsilon_3$ (new noise symbol called 
$\varepsilon_3$ because introduced at control point 3). The concretisation of $\hat z_{[3]}$, using $\varepsilon_1 \in [-1,0]$, is then $\gamma (\hat z_{[3]}) = [0,5]$. 

But it is better to use the constraint on $\varepsilon_1$ to linearize \texttt{z=x*x} at the center of the interval $\varepsilon_1 \in [-1,0]$:
we then write $\hat z_{[3]} = (1.5 + (\varepsilon_1+0.5)) (1.5 + (\varepsilon_1+0.5)) = 2.25 + 1.5+(\varepsilon_1+0.5)+ (\varepsilon_1+0.5)^2$, 
which, using $(\varepsilon_1+0.5)^2 \in [0,0.25]$, can be linearized as 
$\hat z_{[3]} = 3.875 + 3 \varepsilon_1 + 0.125 \varepsilon_3$. Its concretisation is $\gamma (\hat z_{[3]}) = [0.75,4]$.

In the else branch, \texttt{z=x*x} interpreted at control point $5$ with $\varepsilon_1 \in [0,1]$ is linearized by 
$\hat z_{[5]} = (2.5 + (\varepsilon_1-0.5)) (2.5 + (\varepsilon_1-0.5)) = 3.875 + 5 \varepsilon_1 + 0.125 \varepsilon_5$. 
And $\gamma (\hat z_{[5]}) = [3.75,9]$.
\end{example}
\vspace*{-0.2cm}
\subsubsection{Join}
%\label{notphi}

We  need an upper bound operator to combine abstract values coming from different branches. 
The computation of upper bounds (and if possible minimal ones) on constrained affine sets is a difficult task, 
already discussed in several papers~\cite{arxiv08,arxiv09,cav10,nsad12}, and orthogonal to the robustness analysis
presented here. 
We will thus consider we have an upper bound operator on constrained affine sets we note $\sqcup$,  
and focus on the additional term due to discontinuity in tests.

\section{Robustness analysis of finite precision computations}

\label{abstraction}

We introduce here an abstraction which is not only sound in presence of unstable tests, but also exhibits 
the potential discontinuity errors due to these tests. For more concision, we insist here on what is directly linked 
to an accurate treatment of these discontinuities, and rely on previous work~\cite{vmcai11} for the rest. 

\subsection{Abstract values}
\label{abstractvalue}

As in the abstract domain for the analysis of finite precision computations of~\cite{vmcai11}, 
we will see the floating-point computation as a perturbation of a computation in real numbers, and use zonotopic abstractions of 
real computations and errors (introducing respectively noise symbols $\varepsilon_i^r$ and $\varepsilon_j^e$), 
from which we get an abstraction of floating point computations. 
But we make here no assumptions on control flows in tests and will interpret tests independently 
on the real value and the floating-point value. For each branch, we compute conditions for the real 
and floating-point executions to take this branch. The test interpretation
on a zonotopic value~\cite{cav10} lets the affine sets unchanged, but yields constraints on noise symbols. For each branch,
we thus get two sets of constraints:
$\varepsilon^r=(\varepsilon^r_1,\ldots,\varepsilon^r_n) \in \Phi^{X}_r$ for the real control flow (test computed on real values $R^X$), and 
$(\varepsilon^r,\varepsilon^e)=(\varepsilon^r_1,\ldots,\varepsilon^r_n,\varepsilon^e_1,\ldots,\varepsilon^e_m) \in \Phi^{X}_f$ 
for the finite precision control flow (test computed on float values $R^X+E^X$). 

\begin{definition}
 An abstract value $X$, defined at a given control point, for a program with $p$ variables $x_1,\ldots,x_p$, is thus a tuple
$X=(R^X,E^X,D^X,\Phi_r^{X},\Phi_f^{X})$  composed of the following affine sets and constraints, for all $k=1,\ldots,p$:
\[
\left\lbrace
\begin{array}{rllll}
R^X\ : \ \hat r_k^X &=& r_{0,k}^X + \sum_{i=1}^n r_{i,k}^X \, \varepsilon_i^r && \mbox{ where } \varepsilon^r \in \Phi_r^{X}\\
E^X\ : \ \hat e_k^X &=& e_{0,k}^X + \sum_{i=1}^n e_{i,k}^X \, \varepsilon_i^r + \sum_{j=1}^{m} e_{n+j,k}^X \, \varepsilon_j^e && \mbox{ where }  (\varepsilon^r,\varepsilon^e) \in \Phi_f^{X} \\
D^X\ : \ \hat d_k^X &=& d_{0,k}^X + \sum_{i=1}^o d_{i,k}^X \, \varepsilon_i^d &&  \\
 \hat f_k^X &=& \hat r_k^X + \hat{e}_k^X && \mbox{ where }  (\varepsilon^r,\varepsilon^e) \in \Phi_f^{X}
\end{array}
\right.
\]
where  
\begin{itemize}
\item $R^X \in {\cal M}(n+1,p)$ is the affine set defining the real values of variables, and the affine form $\hat r_k^X$ giving the real value of $x_k$, 
is defined on the $\varepsilon_i^r$,
\item $E^X \in {\cal M}(n+m+1,p)$ is the affine set defining the rounding errors (or initial uncertainties) and their propagation through computations as defined in~\cite{vmcai11}, 
and the affine form  $\hat e_k^X$ is defined on the $\varepsilon_i^r$ that model the uncertainty on the real value, and the $\varepsilon_i^e$ that model 
the uncertainty on the rounding errors,
\item $D^X \in {\cal M}(o+1,p)$ is  the affine set defining the discontinuity errors, and $\hat d_k^X$ is defined on noise symbols $\varepsilon_i^d$,
\item the floating-point value is seen as the perturbation by the rounding error of the real value, $\hat{f}_k^X = \hat{r}_k^X + \hat{e}_k^X$.
\item $\Phi_r^X$ is the abstraction of the set of constraints on the  noise symbols such that the real control flow reaches the control point, $\varepsilon^r \in \Phi_r^{X}$,
and $\Phi_f^{X}$ is the abstraction of the set of constraints on the noise symbols such that the finite precision control flow reaches the control point, 
$(\varepsilon^r,\varepsilon^e) \in \Phi_f^{X}$.
\end{itemize}
\end{definition}

A subtlety is that the same affine set $R^X$ is used to define the real value and the floating-point value as a perturbation of the real value, 
but with different constraints: the floating-point value is indeed a perturbation by rounding errors of an idealized computation that would occur with the constraints 
$\Phi_f^{X}$. %We thus compute some real part at each control point even if the real computation, for real constraints $\Phi_r^{X}$, does not reach this point. 

\subsection{Test interpretation}
\label{test}
Consider a test \texttt{e1 op e2}, 
where \texttt{e1} and \texttt{e2} are two arithmetic expressions, and \texttt{op} an operator among $\leq,<,\geq,>,=,\neq$,
the interpretation of this test in our abstract model reduces to the interpretation of \texttt{z op 0}, where \texttt{z}
is the abstraction of expression \texttt{e1 - e2} with affine sets: %Then, using test interpretation 
%over constrained affine sets
\begin{definition}
\label{def-meet1}
Let $X$  be a constrained affine set over $p$ variables. %with $(C^X,E^X) \in {\cal M}(n+1,p) \times {\cal M}(n+m+1,p)$ and $\Phi^X=(\Phi^X_r,\Phi^X_f) \in [-1,1]^n \times [-1,1]^{n+m}$. 
We define $Z = \semb e1 \mbox{ op } e2 \seme X$ by $Y= \semb x_{p+1} := e1 - e2 \seme X$ in 
$Z = drop_{p+1}(\semb x_{p+1} \mbox{ op } 0 \seme Y)$, where function $drop_{p+1}$ returns the affine sets from which component $p+1$ (the intermediary 
variable) has been eliminated.  
\end{definition}

As already said, tests are interpreted independently on the affine sets for real and floating-point value. We use in
Definition \ref{def-meet2}, the test interpretation on constrained affine sets introduced in \cite{cav10}:
\begin{definition}
\label{def-meet2}
Let $X=(R^X,E^X,D^X,\Phi_r^{X},\Phi_f^{X})$ a constrained affine set. % = (R^X,E^X,\Phi^X)$  a constrained affine set with $(R^X,E^X) \in {\cal M}(n+1,p) \times {\cal M}(n+m+1,p)$ 
%and $\Phi^X=(\Phi^X_r,\Phi^X_f) \in [-1,1]^n \times [-1,1]^{n+m}$. 
We define $Z = (\semb x_k \mbox{ op } 0 \seme X$ by 
$$
\left\lbrace
\begin{array}{l}
(R^Z,E^Z,D^Z) = (R^X,E^X,D^X)\\
\Phi^Z_r = \Phi^X_r \bigcap \alpha\left(\varepsilon^r \mid  
r_{0,k}^X + \sum_{i=1}^{n} r_{i,k}^X \varepsilon_i^r \mbox{ op } 0 \right)\\
\Phi^Z_f = \Phi^X_f \bigcap \alpha\left((\varepsilon^r,\varepsilon^e) \mid 
 r_{0,k}^X + e_{0,k}^X + \sum_{i=1}^{n} ( r_{i,k}^X + e_{i,k}^X) \varepsilon_i^r + \sum_{j=1}^{m} e_{n+j,k}^X \varepsilon_j^e  \mbox{ op } 0 \right)
\end{array}
\right.
$$  

\end{definition}

\begin{example}
\label{ex::tests}
Consider the running example. %We now use a superscript $r$ to denote noise symbols for the real value ($\varepsilon_1^r$). 
We start with $\hat r^x_{[1]} = 2 + \varepsilon_1^r$, $\hat e^x_{[1]} = u$. 
The condition for the real control flow to take the \texttt{then} branch is  $\hat r^x_{[1]} = 2 + \varepsilon_1^r \leq 2$, thus $\Phi^r$ is $\varepsilon_1^r \in [-1,0]$.
The condition for the finite precision control flow to take the \texttt{then} branch is $ \hat f^x_{[1]} = \hat{r}^x_{[1]} + \hat{e}^x_{[1]} = 2 + \varepsilon_1^r + u \leq 2$, thus $\Phi^f$ is $\varepsilon_1^r \in [-1,-u]$.
\end{example}

\subsection{Interval concretisation}
\label{sec::int_conc}
The interval concretisation of the value of program variable $x_k$ defined by the abstract value $X=(R^X,E^X,D^X,\Phi_r^{X},\Phi_f^{X})$,
is, with the notations of Section \ref{constrainedzonotopes}:
\[
\left\lbrace
\begin{array}{lll}
 \gamma_r(\hat r_k^X) &=& \Phi^X_r(r_{0,k}^X + \sum_{i=1}^n r_{i,k}^X \, \varepsilon_i^r) \\
\gamma_e(\hat e_k^X) &=& \Phi^X_f(e_{0,k}^X + \sum_{i=1}^n e_{i,k}^X \, \varepsilon_i^r + \sum_{j=1}^m e_{n+j,k}^x \, \varepsilon_j^e)\\
\gamma_d(\hat d_k^X) &=& \Phi^X_f(d_{0,k}^X + \sum_{l=1}^o d_{l,k}^x \, \varepsilon_l^d) \\
\gamma_f(\hat f_k^X) &=& \Phi^X_f(r_{0,k}^X + e_{0,k}^X + \sum_{i=1}^n (r_{i,k}^X + e_{i,k}^X)\, \varepsilon_i^r + \sum_{j=1}^m e_{n+j,k}^x \, \varepsilon_j^e)
\end{array}
\right.
\]

\begin{example}
Consider variable \texttt{y} in the \texttt{else} branch of our running example. 
The interval concretisation of its real value on $\Phi^X_r$, is 
$\gamma_r(\hat r^y_{[4]}) = \Phi^X_r (2 + \varepsilon_1^r) = 2 + [0,1] = [2,3]$. 
The interval concretisation of its floating-point value on  $\Phi^X_f$, is
$\gamma_f(\hat f^y_{[4]}) = \Phi^X_f ( \hat{r}^y_{[4]} + u) = 2 + [-u,1] + u = [2,3+u]$.
Actually, $\hat r^y_{[4]}$ is defined on $\Phi^X_r \cup \Phi^X_f$, as illustrated on Figure \ref{lst::ex1}, because it is both 
used to abstract the real value, or, perturbed by an error term, to abstract the finite precision value.
\end{example}
In other words, the concretisation of the real value is not the same when it actually represents the real value 
at the control point considered ($\gamma_r(\hat r_k^X)$), or when it represents a quantity which will be perturbed to abstract 
the floating-point value (in the computation of $\gamma_f(\hat f_k^X)$). %:
% the conditions on the input values for the execution to go through that control point are not the same for the float or the real 
%control flow, even though they are on some shared noise symbols (the $\varepsilon_i^r$). 

\subsection{Transfer functions: arithmetic expressions}
\label{arithmetic}

We rely here on the transfer functions of \cite{vmcai11} for the full model of values and propagation of errors,
except than some additional care is required due to these constraints. As quickly described in Section \ref{constrainedzonotopes}, constraints on noise symbols 
can be used to refine the abstraction of non affine operations. 
Thus, in order to soundly use the same affine set $R^X$ both for the real value and the floating-point value as a perturbation of a computation in real numbers,
we use constraints $\Phi_r^{X} \cup \Phi_f^{X}$ to abstract transfer functions for the real value $R^X$ in arithmetic expressions. 
Of course, we will then concretize them  either for $\Phi_f^{X}$ or $\Phi_r^{X}$, as described in Section \ref{sec::int_conc}.

\begin{example}
\label{runex3}
Take the running example. In example \ref{runex2}, we computed the real form $\hat r^z$ in both branches, 
interpreting instruction \texttt{z=x*x}, for both sets of constraints $\Phi_r$.
In order to have an abstraction of $\hat r^z$ that can be soundly used both for the floating-point and real values, we will now 
need to compute this abstraction and linearization for $\Phi_r \cup \Phi_f$. In the \texttt{then} branch, 
$\varepsilon_1^r$ is now taken in $[-1,0] \cup [-1,-u] = [-1,0]$, so that $\hat r^z_{[3]} = 3.875+3 \varepsilon_1^r + 0.125 \varepsilon_3^r$ remains unchanged.
But in the \texttt{else} branch, 
$\varepsilon_1^r$ is now taken in $[0,1] \cup [-u,1] = [-u,1]$, so that \texttt{z=x*x} can still be linearized at $\varepsilon_1^r=0.5$ 
but we now have $\hat r^z_{[5]}$ linearized from $(2.5 + (\varepsilon_1^r-0.5)) (2.5 + (\varepsilon_1^r-0.5)) = 6.25 + 5 (\varepsilon_1^r-0.5) + (\varepsilon_1^r-0.5)^2$
where $-0.5-u \leq \varepsilon_1^r-0.5 \leq 0.5$, so that
$\hat r^z_{[5]} = (3.75+\frac{(0.5+u)^2}{2}) + 5 \varepsilon_1^r + \frac{(0.5+u)^2}{2} \varepsilon_5^r = 3.875 + \frac{u+u^2}{2} + 5 \varepsilon_1^r + (0.125+\frac{u+u^2}{2})\varepsilon^r_5$.
\end{example}

\subsection{Join}
\label{join}
In this section, we consider we have upper bound operator $\sqcup$  on constrained affine sets,  
and focus on the additional term due to discontinuity in tests. 
As for the meet operator, we join component-wise the real and floating-point parts.
But, in the same way as for the transfer functions, the join operator depends on the  constraints on the noise symbols:
to compute the affine set abstracting the real value, we must consider 
the join of constraints for real and float control flow, in order to soundly use a perturbation of the real affine set as an 
abstraction of the finite precision value. 

Let us consider the possibility of an unstable test: for a given input, the control flows of the real  
and of the finite precision executions differ. Then, when we join abstract values $X$ and $Y$ coming from the two branches, 
the difference between the floating-point value of $X$ and the real value of $Y$, $(R^X+E^X)-R^Y$, 
and the difference between the floating-point value of $X$ and the real value of $Y$, $(R^Y+E^Y)-R^X$, are also errors due to finite precision. 
The join of errors $E^X$, $E^Y$, $(R^X+E^X)-R^Y$ and $(R^Y+E^Y)-R^X$ can be expressed as 
$E^Z + D^Z$, where $E^Z = E^X \sqcup E^Y$ is the  propagation of classical rounding errors, and 
$D^Z = D^X \sqcup D^Y \sqcup (R^X-R^Y) \sqcup (R^Y-R^X)$ expresses the discontinuity errors. 

The rest of this section will be devoted to an accurate computation of these discontinuity terms. A key point is to use the fact that 
we compute these terms only in the case of unstable tests, which can be expressed as an intersection of 
constraints on the $\varepsilon_i^r$ noise symbols. Indeed this intersection of constraints express the unstable test condition 
as a restriction of the sets of inputs (or equivalently the $\varepsilon_i^r$), such that an unstable test is possible.
 The fact that the same affine set $R^X$ is used both to abstract the real value, and the floating-point 
value when perturbed, is also essential to get accurate bounds.

\begin{definition}
We join two abstract values $X$ and $Y$ by $Z = X \sqcup Y$ defined as $Z=(R^Z,E^Z,D^Z,\Phi^X_r \cup \Phi^Y_r,\Phi^X_f \cup \Phi^Y_f)$ where
$$
\left\lbrace
\begin{array}{l} %{rcl}
(R^Z,\Phi^Z_r \cup \Phi^Z_f)  =  (R^X,\Phi^X_r \cup \Phi^X_f) \sqcup (R^Y,\Phi^Y_r \cup \Phi^Y_f) \\
(E^Z,\Phi^Z_f)  =  (E^X,\Phi^X_f) \sqcup (E^Y,\Phi^Y_f) \\
D^Z  = D^X \sqcup D^Y \sqcup (R^X-R^Y,\Phi^X_f \sqcap \Phi^Y_r) \sqcup 
(R^Y-R^X,\Phi^Y_f \sqcap \Phi^X_r) \\
%\Phi^Z_r = \Phi^X_r \cup  \Phi^Y_r  \\
%\Phi^Z_f = \Phi^X_f \cup  \Phi^Y_f
\end{array}
\right.
$$  
\end{definition}

\begin{example}
Consider again the running example, and let us restrict ourselves for the time being to variable \texttt{y}. 
We join $X=(\hat{r}^y_{[2]}=4+\varepsilon_1^r,\hat{e}^y_{[2]}=u+\delta \varepsilon_2^e,0,\varepsilon_1^r \in [-1,0],(\varepsilon_1^r,\varepsilon_2^e) \in [-1,-u]\times[-1,1])$ coming from 
the \texttt{then} branch with $Y=(\hat{r}^y_{[4]}=2+\varepsilon_1^r,\hat{e}^y_{[4]}=u,0,\varepsilon_1^r \in [0,1],\varepsilon_1^r \in [-u,1])$ coming from the \texttt{else} branch.
Then we can compute the discontinuity error due to the first possible unstable test, when the real takes the \texttt{then} branch and float takes the \texttt{else} branch:
 $\hat{r}^y_{[4]}-\hat{r}^y_{[2]} = 2+\varepsilon_1^r - 4+\varepsilon_1^r = -2$, 
for $\varepsilon_1^r \in \Phi^Y_f \cap \Phi^X_r = [-u,1] \cap [-1,0] = [-u,0]$ (note that the restriction on $\varepsilon_1^r$ is not used here but will be in more general cases).
The other possibility of an unstable test, when the real takes the \texttt{else} branch and float takes the \texttt{then} branch, occurs 
for $\varepsilon_1^r \in \Phi^X_f \cap \Phi^Y_r = [-1,-u] \cap [0,1] = \emptyset$: the set of inputs for which  this unstable test can occur is empty, it never occurs.
 We get $Z=(3+\varepsilon_6^r,u+\delta \varepsilon_2^e,-2\chi_{[-u,0]}(\varepsilon_1^r),(\varepsilon_1^r,\varepsilon_6^r) \in [-1,1]^2,(\varepsilon_1^r,\varepsilon_6^r,\varepsilon_2^e) \in [-1,1]^3)$. %Note that for joining the real values $r^y_{[2]}$ and $r^y_{[4]}$, we do not use the argmin formula as introduced in Section \ref{notphi}, because 
% $r_{[3]}^y = 3 + \varepsilon_6^r$, with a new noise symbol $\varepsilon_6^r$, is a tighter upper bound here: but these considerations are not 
%at the heart of this paper.
\end{example}

\section{Technical matters}
\label{sec::technical}

We gave here the large picture. Still, there are some technical matters to consider in order to efficiently compute accurate bounds for the discontinuity error 
in the general case. We tackle some of them in this section. 

\subsection{Constraint solving using slack variables}

Take the following program, where the real value of inputs $x$ and $y$ are in range [-1,1], and  both have 
an error bounded in absolute value by some small value $u$: 
\begin{lstlisting}[frame=single,language=C,escapechar=@,basicstyle=\scriptsize] %,caption={Running example},label=lst::ex1,basicstyle=\tiny]

x := [@-@1,1] + [@-@u,u]; // [1] ; 0 < u << 1
y := [@-@1,1] + [@-@u,u]; // [2]  
if (x < y)
   t = y @-@ x;         // [3]
else
   t = x @-@ y;         // [4]
\end{lstlisting}
The test can be unstable, we want to prove the treatment continuous. 
Before the test, $\hat{r}^x_{[1]} = \varepsilon^r_1$, $\hat{e}^x_{[1]} = u \varepsilon^e_1$, $\hat{r}^y_{[2]} = \varepsilon^r_2$, $\hat{e}^y_{[2]} = u \varepsilon^e_2$.
The conditions for the control flow to take the \texttt{then} branch are
$\varepsilon^r_1<\varepsilon^r_2$ for the real execution, and $\varepsilon^r_1 + u \varepsilon^e_1 <\varepsilon^r_2 + u \varepsilon^e_2$ for the float execution.
The real value of $t$ in this branch is $\hat{r}^t_{[3]} = \varepsilon^r_2 - \varepsilon^r_1$. %, $e^t_{[3]} = u \varepsilon^e_2 - u \varepsilon^e_1 + 2 u \varepsilon^e_3$, 
%where we added a new error term to account for the rounding error of $y-x$. 
In the \texttt{else} branch, the conditions are the reverse and $\hat{r}^t_{[4]} = \varepsilon^r_1 - \varepsilon^r_2$. %, $e^t_{[4]} = u \varepsilon^e_1 - u \varepsilon^e_2 + 2 u \varepsilon^e_4$.

Let us consider the possibility of unstable tests. % and the corresponding discontinuity errors. 
The conditions for the floating-point to take the else branch while the real takes the then branch are 
$\varepsilon^r_1 + u \varepsilon^e_1 \geq \varepsilon^r_2 + u \varepsilon^e_2$ and $\varepsilon^r_1<\varepsilon^r_2$, 
from which we can deduce $-2 u < \varepsilon^r_1-\varepsilon^r_2 < 0$.
Under these conditions, we can bound $\hat{r}^t_{[4]} - \hat{r}^t_{[3]} = 2 (\varepsilon^r_1 - \varepsilon^r_2)  \in [-4u,0]$. The other 
 unstable test is symmetric, we thus have proven that the discontinuity error is 
of the order of the error on inputs, that is the conditional block is robust.

Note that on this example, we needed more than interval constraints on noise symbols, and would in general have to solve linear programs. 
However, we can remark that constraints on real and floating-point parts share the same subexpressions on the $\varepsilon^r$ noise symbols. 
Thus, introducing slack symbols such that the test conditions are expressed on these slack variables, we can keep the full precision 
when solving the constraints in
intervals. Here, introducing $\varepsilon_3^r = \varepsilon^r_1-\varepsilon^r_2$, the unstable test condition is expressed 
as $\varepsilon_3^r < 0$ and $\varepsilon_3^r > -2u$. 
This is akin to using the first step of the simplex method for linear programs, where slack
variables are introduced to put the problem in standard form.

\subsection{Linearization of non affine computations near the test condition}
There can be a need for more accuracy near the test conditions: one situation is when we have successive joins, where several 
tests may be unstable, such as the interpolator example presented in the experiments. In this case, it is necessary to keep some information on the states
at the extremities when joining values (and get rid of this extra information as soon as we exit the conditional block).  
More interesting, there is a need for more accuracy near the test condition when the conditional block contains some non linear computations. 

\begin{example}
\label{ex3}
Consider again the running example. We are now interested in variable $z$. 
There is obviously no discontinuity around the test condition; 
still, our present abstraction is not accurate enough to prove so.
Remember from Examples \ref{runex2} and \ref{runex3} that we linearize in each branch \texttt{x*x} for $\Phi_r \cup \Phi_f$,
introducing new noise symbols  $\varepsilon_3^r$ and  $\varepsilon_5^r$.
%$\varepsilon_1^r \in [-u,1]=[-u,1]$, and \texttt{z=x*x} is linearized to give
%$\hat r^z_{[5]} = 3.875 + \frac{u+u^2}{2} + 5 \varepsilon_1^r + (0.125+\frac{u+u^2}{2})\varepsilon^r_5$.
Let us consider the unstable test when the real execution takes the then branch and the floating-point execution the other branch,
the corresponding discontinuity error $\hat r^z_{[5]} - \hat{r}^z_{[3]}$, under unstable test constraint $ -u < \varepsilon_1^r < 0$, is:
\begin{equation}
\label{eq1}
 \hat{r}^z_{[5]} - \hat{r}^z_{[3]} = \frac{u+u^2}{2} + 2 \varepsilon_1^r  + (0.125+\frac{u+u^2}{2})\varepsilon^r_5 -  0.125 \varepsilon_3^r.  
\end{equation}
In this expression, from constraint $ -u < \varepsilon_1^r < 0$ we can prove that $\frac{u+u^2}{2} + 2 \varepsilon_1^r  + \frac{u+u^2}{2}\varepsilon^r_5$
is of the order of the input error $u$. But the new noise term $0.125(\varepsilon^r_5-\varepsilon_3^r)$ is only bounded by $[-0.25,0.25]$.
We thus cannot prove continuity here. This is illustrated on the left-hand side of Figure \ref{fig::linearization}, on which we represented the zonotopic 
abstractions $\hat r^z_{[3]}$ and $\hat  r^z_{[5]}$: it 
clearly appears that the zonotopic abstraction is not sufficient to accurately bound the discontinuity error (in the ellipse),
that will locally involve some interval-like computation.
Indeed, in the linearization of $\hat r^z_{[3]}$ (resp $\hat r^z_{[5]}$), we lost the correlation between the new symbol
$\varepsilon_3^r$ (resp $\varepsilon_5^r$), and symbol $\varepsilon_1^r$ on which the unstable test constraint is expressed. 
As a matter of fact, we can locally derive in a systematic way some affine bounds for the new noise symbols used for linearization 
in terms of the existing noise symbols, using the interval affine forms
of \cite{SAS07}, centered at the extremities of the constraints $(\Phi_r^X \cup \Phi_f^X)(\varepsilon_i^r)$ of interest. 

In the \texttt{then} branch, we have $\varepsilon_1^r \in [-1,0]$, and \texttt{z=x*x} is linearized 
from $3.75 + (\varepsilon_1^r+0.5)+ (\varepsilon_1^r+0.5)^2$, 
 using $(\varepsilon_1+0.5)^2 \in [0,0.25]$, into $\hat r^z_{[3]} = 3.875+3 \varepsilon_1^r + 0.125 \varepsilon_3^r$.
%Similarly in the \texttt{else} branch. 
We thus know at linearization time that $\varepsilon_3^r = f(\varepsilon_1^r) = 8(\varepsilon_1^r+0.5)^2 - 1$.
Using the mean value theorem around $\varepsilon_1^r=0$ and restricting $\varepsilon_1^r \in [-0.25,0]$, we write
$$\varepsilon_3^r(\varepsilon_1^r) = f(0) + \Delta \varepsilon_1^r,$$ where interval $\Delta$
bounds the derivative $f'(\varepsilon_1^r)$ in the range $[-0.25,0]$. We get
$\varepsilon_3^r = 1 + 16 ([-0.25,0]+0.5) \varepsilon_1^r =  1+[4,8]\varepsilon_1^r,$
which we can also write
$1+8\varepsilon_1^r \leq \varepsilon_3^r \leq 1+4\varepsilon_1^r$ for $\varepsilon_1^r \in [-0.25,0]$.
Variable \texttt{z} can thus locally (for $\varepsilon_1^r \in [-0.25,0]$) be expressed more accurately as a function of 
$\varepsilon_1^r$, this is what is represented by the darker triangular region inside the zonotopic abstraction, 
on the right-hand side of Figure \ref{fig::linearization}. \\
%\vspace*{-0.3cm}
\begin{figure}
\centering
\begin{tabular}{ll}
\begin{tikzpicture}[xscale=5,yscale=0.5]

\draw[xstep=0.5cm,ystep=1cm,gray,very thin] (1.5,2) grid (2.5,7); 
\draw[->] (1.5,2) -- (2.7,2) node[anchor=south] {$\varepsilon_1^r$};
%\draw (1 cm,0 cm)  node[below] {-1};
\draw (1.5 cm,2 cm)  node[below] {-0.5};
\draw (2 cm,2 cm)  node[below] {0};
\draw (2.5 cm,2 cm)  node[below] {0.5};
%\draw (3 cm,0 cm)  node[below] {1};
\draw[->] (1.5,2) -- (1.5,8) node[anchor=east] {$z$};
\foreach \y/\ytext in {2,3,4,5,6,7} 
\draw (1.5 cm,\y cm) node[left] {$\ytext$};
%\only<1>{\filldraw[opacity=0.5, fill=yellow!30!white, draw=black] (-10,0) -- (50,0) -- (90,10) -- (-10,10) -- cycle;

%\filldraw[opacity=0.5, fill=green!30!white, draw=green!50!black] (1,0.750) -- (1,1) --(2,4) -- (2,3.750) -- cycle;
\filldraw[opacity=0.5, fill=green!30!white, draw=green!50!black] (1.5,2.25) -- (1.5,2.5) --(2,4) -- (2,3.750) -- cycle;

\filldraw[opacity=0.5, fill=blue!30!white, draw=blue!50!black] (1.95,3.55) -- (1.95,3.8) -- (2.5,6.4) -- (2.5,6.15) -- cycle;

\draw [domain=1.5:2.5,purple,thick,dashed] plot (\x,{\x*\x});

\draw[red,dashed] (1.95,2) --  (1.95,7);
\draw[red,dashed] (2,2) --  (2,7); 

\draw[red] (1.975,3.75) ellipse (0.2 and 1);
%\draw[red] (2. cm, -5 cm) node[anchor=south] {$ -u < \varepsilon_1^r \leq 0$};

\draw[green] (1.6 cm, 2.8 cm) node[anchor=south] {$r^z_{[3]}$};
\draw[blue] (2.1 cm, 4.8 cm) node[anchor=south] {$r^z_{[5]}$};
\end{tikzpicture}
&
\begin{tikzpicture}[xscale=15,yscale=2]

\draw[xstep=0.125cm,ystep=0.5cm,gray,very thin] (1.75,3) grid (2.,4); 
\draw[->] (1.75,3) -- (2.05,3) node[anchor=south] {$\varepsilon_1^r$};

\draw (1.75 cm,3 cm)  node[below] {-0.25};
\draw (2. cm,3 cm)  node[below] {0};
\draw[->] (1.75,3) -- (1.75,4.2) node[anchor=east] {$z$};
\foreach \y/\ytext in {3,4} 
\draw (1.75 cm,\y cm) node[left] {$\ytext$};

\filldraw[opacity=0.5, fill=green!30!white, draw=green!50!black] (1.75,3.0) -- (1.75,3.25) --(2,4) -- (2,3.750) -- cycle;
\draw [domain=1.75:2.,purple,thick,dashed] plot (\x,{\x*\x});
\filldraw[opacity=0.5, fill=brown!50!white, draw=black] (1.75,3) -- (1.75,3.125) --(2,4) -- cycle;
\end{tikzpicture}
\end{tabular}
\caption{Improvement by local linearization for non affine computations}
\label{fig::linearization}
\end{figure}
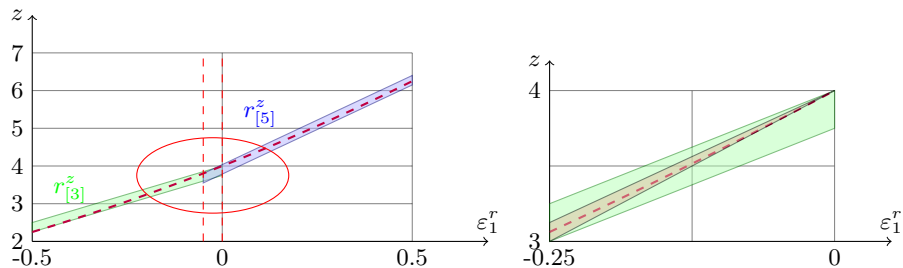
%\vspace*{-0.3cm}
In the same way, $\varepsilon_5^r$ can be expressed in the \texttt{else} branch as an affine form $1+\Delta' \varepsilon_1^r$ with interval 
coefficient $\Delta'$, so that with the unstable test constraint $ -u < \varepsilon_1^r < 0$, we can deduce 
from Equation (\ref{eq1}) that there exists some constant $K$ such that $|\hat{r}^z_{[5]} -\hat{r}^z_{[3]}| \leq K u$, that is the test is robust.
Of course, we could refine even more the bounds for the discontinuity error by considering linearization on smaller intervals around the boundary condition.
\end{example}

%%%%%%%%%%%%%%%%%%%%%%%%%%%%%%%%%%%%%%%%%%%%%%%%%%%%%%%%%%%%%%%%%%%%%%%%%%%%%%%%%%%%%%%%%%%%%%%%%%%%%%%%%%%%
\section{Experiments}

\label{experiments}

In what follows, we analyze some examples inspired by industrial codes and literature,
with our implementation in our static analyzer FLUCTUAT. 

\paragraph{A simple interpolator}

%interpol2.c but also more complicated ones...(\texttt{interpolateur.c}).

The following example implements %is based on an industrial example; it is an implementation of
an interpolator, affine by sub-intervals, as classically found in critical
embedded software. It is a robust implementation indeed. In the code below, we used
the FLUCTUAT assertion 
\texttt{FREAL\_WITH\_ERROR(a,b,c,d)}
to denote an abstract value (of resulting type \texttt{float}), whose corresponding real values are 
$x \in [a,b]$, and whose corresponding floating-point values are of the form $x+e$, with $e \in [c,d]$. 
%typedef struct 
%{
%  float x;
%  float y;
%} paire;
\begin{lstlisting}[language=C,frame=single,escapechar=@,basicstyle=\tiny] %caption={A simple interpolator},label=lst::interpolator,basicstyle=\tiny]

  float R1[3], E, res;
  R1[0] = 0;  R1[1] = 5 * 2.25; R1[2] = R1[1] + 20 * 1.1;
  E = FREAL_WITH_ERROR(0.0,100.0,@-@0.00001,0.00001);
  if (E < 5)
    res = E*2.25 + R1[0];
  else if (E < 25)
    res = (E@-@5)*1.1 + R1[1];
  else
    res = R1[2];
  return res; 
\end{lstlisting}

The analysis finds that the interpolated \texttt{res}
is within [-2.25e-5,33.25], with an error within [-3.55e-5,2.4e-5], that is of the order of magnitude of the input error despite unstable tests.

\paragraph{A simple square root function} % \texttt{newnewsqrt.c}}

This example is a rewrite in some particular case, of an actual implementation of a square root
function, in an industrial context: 
\begin{lstlisting}[language=C,frame=single,escapechar=@,basicstyle=\tiny]%,caption={A simple square root},label=lst::squareroot,basicstyle=\tiny]

  double sqrt2 = 1.414213538169860839843750;
  double S, I;  I = DREAL_WITH_ERROR(1,2,0,0.001);
  if (I>=2) 
    S = sqrt2*(1+(I/2@-@1)*(.5@-@0.125*(I/2@-@1)));
   else 
    S = 1+(I@-@1)*(.5+(I@-@1)*(@-@.125+(I@-@1)*.0625)); 
\end{lstlisting}
With the former type of analysis within FLUCTUAT, we get the unsound
result - but an unstable test is signalled - 
that \texttt{S} is proven in the real number semantics to be in [1,1.4531] with
a global error in [-0.0005312,0.00008592].
%\end{itemize}

As a matter of fact, the function does not exhibit a big discontinuity, but still, it
is bigger than the one computed above. At value 2, 
the function in the \texttt{then} branch computes \texttt{sqrt2} which is approximately
1.4142, whereas the \texttt{else} branch computes 1+0.5-0.125+0.0625=1.4375. Therefore,
for instance, for a real number input of 2, and a floating-point number input of 2+$ulp(2)$,
we get a computation error on $S$ of the order of 0.0233.
FLUCTUAT, using the domain described in this paper finds
that \texttt{S} is in the real number semantics within [1,1.4531] with
a global error within [-0.03941,0.03895], the discontinuity at the test accounting
for most of it, i.e. an error within [-0.03898,0.03898] (which is coherent with respect
to the rough estimate of 0.0233 we made).

\paragraph{Transmission shift from \cite{DBLP:conf/rtss/MajumdarS09}}

We consider here the program from \cite{DBLP:conf/rtss/MajumdarS09} 
that implements a simple model of a transmission shift: according to a variable \texttt{angle} measured, and
the \texttt{speed}, lookup tables are used to compute \texttt{pressure1} and \texttt{pressure2}, and
deduce also the current \texttt{gear} (3 or 4 here).
As noted in \cite{DBLP:conf/rtss/MajumdarS09}, \texttt{pressure1} is robust.
But a small deviation in \texttt{speed} can cause a large deviation in the output
\texttt{pressure2}. 
As an example, when \texttt{angle} is 34 and \texttt{speed} is 14, 
\texttt{pressure2} is 1000. But if there is an error of 1 in the measurement of
\texttt{angle}, so that its value is 35 instead of 34, then 
\texttt{pressure2} is found to be 0. Similarly with an error of 1 on 
\texttt{speed}: if it is wrongly measured to be 13 instead of 14, \texttt{pressure2}
is found equal to 0 instead of 1000, again. 

This is witnessed by our discontinuity analysis. 
For \texttt{angle} in [0,90], with an error in [-1,1] and \texttt{speed}
in [0,40], with an error in [-1,1], we find \texttt{pressure1}
equal to 1000 without error
and \texttt{pressure2} in [0,1000] with an error in [-1000,1000], mostly
due to test \texttt{if (oval <= 3)} in function \texttt{lookup2\_2d}. The treatment on \texttt{gear}
is found discontinuous, because of test \texttt{if (3*speed <= val1)}.

\begin{figure}[htbp]
\begin{center}
\epsfig{file=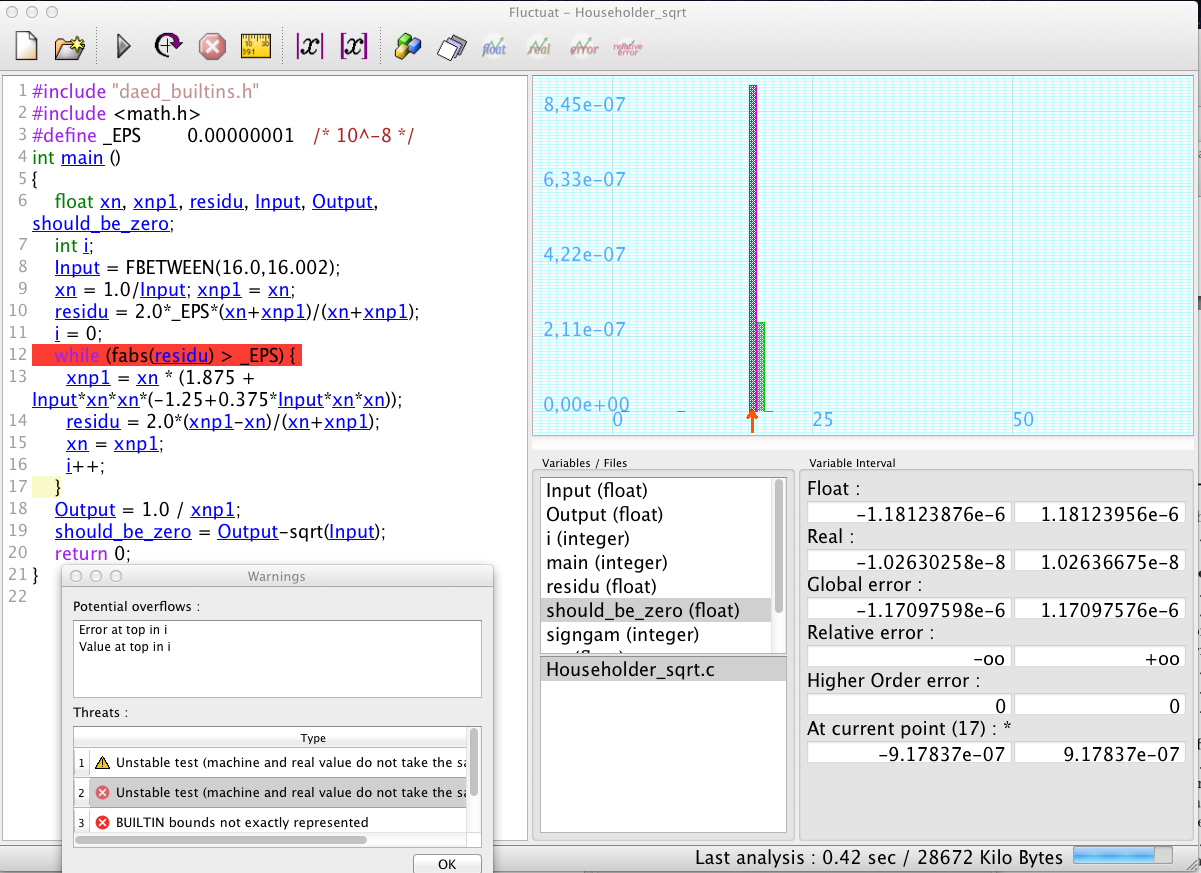,width=12.cm,height=6cm,clip=}
\caption{Fluctuat analysis of the Householder scheme: error due to unstable test is purple}
\label{fig::householder}
\end{center}
\end{figure}

\paragraph{Householder}
Let us consider the C code printed on the left hand side of Figure \ref{fig::householder}, which presents the results of the 
analysis of this program by FLUCTUAT. 
This program computes in variable \texttt{Output}, an approximation of the square root 
of variable \texttt{Input}, which is given here in a small interval [16.0,16.002]. 
The program iterates a polynomial approximation until the difference 
between two successive iterates \texttt{xn} and \texttt{xnp1} is smaller than some stopping criterion.
At the end, it checks that something indeed close to the mathematical square root is computed, by adding 
instruction \texttt{should\_be\_zero = Output-sqrt(Input);}
 Figure \ref{fig::householder} presents the result of the analysis for the selected variable \texttt{should\_be\_zero}, at the end of the program. 
The analyzer issues an unstable test warning, which line in the program is highlighted in red. 
On the right hand side, bounds for the floating-point, real values and error of \texttt{should\_be\_zero} are
printed. The graph with the error bars represents the decomposition on the error on its provenance on the lines of the 
program analyzed: in green are standard rounding errors, in purple the discontinuity error due to unstable tests. When 
an error bar is selected (here, the purple one), the bounds for this error are printed in the boxes denoted ``At current point''. 
%\vspace*{-0.3cm}
%\vspace*{-0.3cm}
The analyzer here proves that when the program terminates, the difference in real numbers 
between the output and the mathematical square root of the input is bounded by $[-1.03e^{-8},1.03e^{-8}]$: the 
algorithm in real numbers indeed computes something close to a square root, and the method error is of the order
of the stopping criterion \texttt{eps}. The floating-point value of the difference is only bounded in $[-1.19e^{-6},1.19e^{-6}]$, and the error
mainly comes from the instability of the loop condition: this signals a difficulty of this scheme when executed in simple precision. And indeed, 
this scheme converges very quickly in real numbers (FLUCTUAT proves that it always converges in 6 iterations for the given range of inputs), 
but there exists input values in  [16.0,16.002] for which the floating-point program never converges.

\section{Conclusion}

We have proposed an abstract interpretation based static analysis of the robustness of finite precision implementations, as a generalization 
of both software robustness or continuity analysis and finite precision error analysis, by abstracting the impact of finite precision in 
numerical computations and control flow divergences. We have demonstrated its accuracy, although it could still be improved. We could also possibly use 
this abstraction to automatically generate inputs and parameters leading to instabilities. 
In all cases, this probably involves resorting to more sophisticated constraint 
solving: indeed our analysis can generate constraints on noise symbols, which we only partially use for the time being. We would thus like to go 
along the lines of \cite{rueher2012}, which refined the results of a previous version of FLUCTUAT using constraint solving, but using 
more refined interactions in the context of the present abstractions.

\bibliographystyle{plain} 
\bibliography{Discontinuity13}

\end{document}